\documentclass{iopart}
\usepackage[dvips]{graphicx}
\usepackage{iopams}
\usepackage{setstack}

\usepackage{url}
\newcommand{\abs}[1]{\left|#1\right|}
\newcommand{\un}[1]{\mathrm{\,#1}}
\newcommand{\mc}[1]{\mathcal{#1}}
\newcommand{\wt}[1]{\widetilde{#1}}
\newcommand{\F}{\mathcal{F}}

\def\mbf#1{\ensuremath{\mathchoice{\mbox{\boldmath$\displaystyle#1$}}
{\mbox{\boldmath$\textstyle#1$}}
{\mbox{\boldmath$\scriptstyle#1$}}
{\mbox{\boldmath$\scriptscriptstyle#1$}}}}

\def\vect#1{\vec{#1}}
\usepackage{tensor}
\def\tens#1{\tensor{#1}}

\def\expected#1{E\left[ #1 \right]}

\def\tauref{\tau_{\mathrm{ref}}}
\def\synthLISA{\mathrm{synthLISA}}
\def\LISAsim{\mathrm{LISAsim}}
\def\O{\mc{O}}
\def\SFT{\mathrm{SFT}}
\def\doppler{\theta}
\def\lon{\lambda}
\def\lat{\beta}
\newcommand{\skypos}{{\{\lat, \lon\}}}

\def\eps{\epsilon}
\def\epsA{\eps_{\A}}

\def\key{\mathrm{key}}
\def\cand{\mathrm{cand}}
\def\A{\mc{A}}
\def\M{\mc{M}}
\def\mHz{\mathrm{mHz}}

\def\dcc{LIGO-P070029-01-Z}

\begin{document}
%\draft
\title[$\F$-statistic search for WD binaries in the first MLDC]
{$\F$-statistic search for white-dwarf binaries in the first Mock LISA Data Challenge}
\author{Reinhard Prix and John T Whelan}
\address{Max-Planck-Institut f\"{u}r Gravitationsphysik
  (Albert-Einstein-Institut), D-14476 Potsdam, Germany} 
\begin{abstract}
  The $\F$-statistic is an optimal detection statistic for
  continuous gravitational waves, i.e., long-duration
  (quasi-)monochromatic signals with slowly-varying intrinsic frequency.
  This method was originally developed in the context of ground-based
  detectors, but it is equally applicable to LISA where many signals
  fall into this class of signals.
  We report on the application of a LIGO/GEO $\F$-statistic
  code to LISA data-analysis using the long-wavelength limit (LWL),
  and we present results of our search for white-dwarf binary signals
  in the first Mock LISA Data Challenge.  
  Somewhat surprisingly, the LWL is found to be sufficient -- even at
  high frequencies -- for \emph{detection} 
  of signals and their accurate localization on the sky and in
  frequency, while a more accurate modelling of the TDI response only seems
  necessary to correctly estimate the four amplitude parameters. 
\end{abstract}
\ead{reinhard.prix@aei.mpg.de, john.whelan@aei.mpg.de}
%\maketitle

\section{Introduction}
\label{s:intro}

The Mock LISA Data Challenge (MLDC) \cite{mldc:_homepage} has the
purpose of encouraging the development of LISA data-analysis tools and
assessing the technical readiness of the community to perform
gravitational-wave (GW) astronomy with LISA.  
The first round of the MLDC was released in June 2006
\cite{2006gr.qc.....9105A}, the submission deadline was in
December 2006 and a report summarizing the submitted results has been
published \cite{GWDAW_MLDC1}. 
The challenges consisted of several data-sets containing different
types of simulated sources and LISA noise. The three types of sources
are white-dwarf binary signals (WD), coalescing supermassive black
holes (SMBHs) and extreme mass-ratio inspirals (EMRIs). 

The data analysis of LISA poses a few specific difficulties not
encountered in ground-based detectors: the signal
(reduced) wavelength is typically not long compared to the arm-length of the
detector, so the long-wavelength limit (LWL) does not generally apply. 
Furthermore, in order to cancel the dominating laser-frequency
noise, one has to analyze intricate algebraic combinations of
time-delays between spacecraft instead of simple ``strain'', an
approach known as \emph{time-delay interferometry} (TDI). 
Another difficulty stems from the large number of detectable sources
in the LISA bandwidth, which complicates their separate detection and
parameter estimation, usually referred to as the ``confusion problem''.

Most of the relevant signals in LISA (WD, SMBH, EMRI) will be long-lasting
(of the order of a year) and are \mbox{(quasi-)}monochromatic with
slowly-varying intrinsic frequency $f(\tau)$; in this sense they
belong to the class of \emph{continuous GWs}.
In the case of ground-based detectors the typical sources of
continuous GWs are spinning neutron stars with non-axisymmetric
deformations.    
One of the standard tools developed for these searches is the
$\F$-statistic \cite{jks98:_data}, which is an optimal detection
statistic (in the sense of the Neyman-Pearson lemma) based on matched
filtering.   
We have restricted our searches in the first MLDC to WD-binary
signals, which are very similar to GWs from spinning neutron stars,
which have very little intrinsic frequency evolution $\dot{f}$ (in fact,
here it was $\dot{f}=0$) and constant orientation and polarization.  
Contrary to the approach used in \cite{krolak04:_optim_lisa,krolak07:_mldc}, 
we use an $\F$-statistic code developed for the continuous-wave search
in LIGO/GEO, with only minimal modifications to adapt it to LISA.  
In particular, we use the LWL at all frequencies, which turns out to
work surprisingly well even at high frequencies where the reduced wavelength
is comparable to the LISA arm length.

\section{Methods and Pipeline}

\subsection{Continuous Gravitational Wave Signals}

A system with an oscillating mass quadrupole moment
emits GWs described, far from the source, by the metric perturbation 
\begin{equation}
  \tens{h} = A_{+} \cos(\phi_0 + \phi)\, \tens{e}_{\!+}
  +  A_{\times} \sin(\phi_0 + \phi)\, \tens{e}_{\!\times}\,,
\end{equation}
where
$\tens{e}_{\!+} =\vect{e}_x \otimes \vect{e}_x - \vect{e}_y \otimes \vect{e}_y$
and
$\tens{e}_{\!\times} = \vect{e}_x \otimes \vect{e}_y + \vect{e}_y \otimes \vect{e}_x$ 
are
the polarization basis tensors constructed from a right-handed basis
$\{\vect{e}_x,\,\vect{e}_y,\,\vect{e}_z\}$ with $\vect{e}_z$ pointing
in the direction of propagation of the wave, described by the ecliptic
latitude $\lat$ and longitude $\lon$, and $\vect{e}_x$ and $\vect{e}_y$
along the principal polarization axes.
In an inertial reference frame, such as the solar-system barycenter
(SSB), the phase of this (quasi-)periodic signal can be
written as $\phi(\tau) = 2\pi\int_{\tauref}^\tau f(\tau')\,d \tau'$,
in terms of the (slowly-varying) intrinsic GW frequency 
$f(\tau) = f(\tauref) + \dot{f}(\tauref)\,\Delta\tau + \ldots\,$,
where $\tauref$ is a reference time at which the frequency and
spindown parameters are defined, and $\Delta\tau \equiv \tau - \tauref$.
The WD signals in the first MLDC were restricted to have a constant
intrinsic frequency, i.e., $f(\tau) = f$. This is a realistic
assumption at low frequencies $f\sim 1\,\mHz$, but at higher
frequencies $f \sim 10\,\mHz$ one would probably have to include one
derivative $\dot{f}$ (e.g.\ see \cite{krolak04:_optim_lisa}) in an
actual search on LISA data.
In the case of a binary system for which
orbital evolution due to GW emission can be neglected, the principal
polarization axes are found by taking the unit vector $\vect{e}_x$ to lie
in the orbital plane and $\vect{e}_y$ in the hemisphere containing the
orbital angular momentum. The polarization amplitudes are
$A_{+}=h_0(1+\cos^2\iota)/2$ and $A_{\times}=h_0\cos\iota$, where
$h_0$ is usually referred to as the \emph{amplitude} of the GW, and
$\iota$ is the inclination angle between the propagation direction
$\vect{e}_z$ and the normal to the orbital plane. 
In order to separate the sky position $\skypos$ from the source
polarization, it is useful to consider a polarization basis associated
only with the sky position; this is done by defining a right-handed
orthonormal basis 
$\{\vect{e}_\xi,\,\vect{e}_\eta,\,\vect{e}_\zeta\}$ with
$\vect{e}_\zeta = \vect{e}_z$ as the propagation direction, 
$\vect{e}_\xi$ lying in the ecliptic plane and $\vect{e}_\eta$ in the
northern hemisphere.  The alternative polarization basis is then
$\tens{\varepsilon}_{\!+} = \vect{e}_\xi \otimes \vect{e}_\xi - \vect{e}_\eta \otimes \vect{e}_\eta$
and
$\tens{\varepsilon}_{\!\times} = \vect{e}_\xi \otimes \vect{e}_\eta + \vect{e}_\eta \otimes \vect{e}_\xi$,
and the principal polarization axes of the GW are determined by the
angle $\psi$ from $\vect{e}_\xi$ to $\vect{e}_x$, measured
counter-clockwise around $\vect{e}_z = \vect{e}_\zeta$, i.e.,  
\begin{equation}
  \eqalign{
    \tens{e}_{\!+} &= \;\;\; \tens{\varepsilon}_{\!+} \,\cos 2\psi +
    \tens{\varepsilon}_{\!\times} \, \sin 2\psi\,,\\
    \tens{e}_{\!\times} &= - \tens{\varepsilon}_{\!+} \,\sin 2\psi +
    \tens{\varepsilon}_{\!\times} \, \cos 2\psi\,.}
  \label{eq:6}
\end{equation}
In terms of this alternative polarization basis, the GW tensor can
be written as 
\begin{equation}
  \label{eq:4}
  \tens{h}(\tau) = \sum_{\mu=1}^4 \mc{A}^\mu\, \tens{h}_{\mu}(\tau)\,,
\end{equation}
where the four \emph{amplitude parameters} $\{\mc{A}^\mu\}$ are  
\begin{equation}
  \eqalign{
    \mc{A}^1 = A_{+}\cos\phi_0\cos 2\psi - A_{\times}\sin\phi_0\sin 2\psi\,,\\
    \mc{A}^2 = A_{+}\cos\phi_0\sin 2\psi + A_{\times}\sin\phi_0\cos 2\psi\,,\\
    \mc{A}^3 = - A_{+}\sin\phi_0\cos 2\psi - A_{\times}\cos\phi_0\sin 2\psi\,,\\
    \mc{A}^4 = - A_{+}\sin\phi_0\sin 2\psi + A_{\times}\cos\phi_0\cos 2\psi\,,}
  \label{eq:5}
\end{equation}
while the tensors $\{\tens{h}_{\mu}\}$ depend on the frequency $f(\tau)$ and
the sky position $\skypos$, namely 
\begin{equation}
  \label{eq:7}
  \eqalign{
    \tens{h}_{1}(\tau) = \tens{\varepsilon}_{\!+} \, \cos\phi(\tau)\,,
    &\qquad
    \tens{h}_{2}(\tau) = \tens{\varepsilon}_{\!\times}\, \cos\phi(\tau)\,,
    \\
    \tens{h}_{3}(\tau) = \tens{\varepsilon}_{\!+} \, \sin\phi(\tau)\,,
    &\qquad
    \tens{h}_{4}(\tau) = \tens{\varepsilon}_{\!\times}\, \sin\phi(\tau)\,.
    }
\end{equation}
Note that the geometrical conventions for the amplitude parameters used
here are consistent with the LIGO/GEO conventions for continuous GWs (e.g.\
\cite{lsc06:_coher_scorp_x}), but differ from the LISA/MLDC
conventions \cite{GWDAW_MLDC1,2007gr.qc.....1170A}. The translation
into MLDC conventions is given by: 
``Amplitude'' $\equiv h_0/2$, $\iota \rightarrow \pi - \iota$,
$\psi \rightarrow \pi/2 - \psi$ and $\phi_0 \rightarrow \phi_0 + \pi$.

\subsection{LISA Response in the Long-Wavelength Limit}

\begin{figure}
  \centering
  \includegraphics[width=0.5\textwidth]{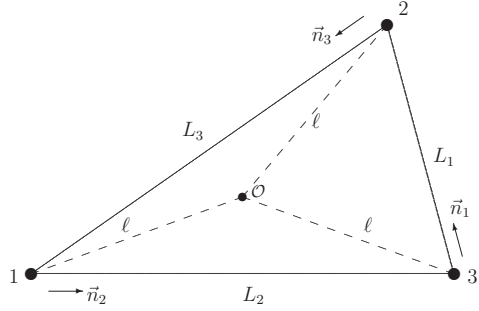}
  \caption{LISA configuration and TDI conventions used.}
  \label{f:LISA}
\end{figure}

The LISA design consists of three spacecraft with laser links between
each pair, in a geometry illustrated in \fref{f:LISA}.
The MLDC data were generated by two different programs:  Synthetic
LISA \cite{synthLISA} simulates a detector output consisting of
Doppler shifts of the LISA lasers due to relative motion of the
spacecraft, while LISA Simulator \cite{LISAsim} simulates the phase
differences between laser light following different paths between the
spacecraft. In both cases the  underlying variables are combined with
appropriate time shifts to form TDI observables which cancel the
(otherwise dominating) laser frequency noise \cite{TDI:_1999,TDI:_2004,krolak04:_optim_lisa}.
One choice of such TDI quantities is the set of three observables
$\{X, Y, Z\}$, which were used to publish the data of the first MLDC.
These observables, which can be thought of as representing the output
of three virtual ``detectors'' $I$, are related to the gravitational wave
$\tens{h}$ through somewhat involved expressions depending on the
frequency and propagation direction of the wave.  
However, in the LWL approximation, in which the reduced wavelength 
$c/(2\pi f)$ is assumed to be large compared to the distance between
the spacecraft, i.e., $f \ll 10\,\mHz$, the responses can be
approximated (assuming $L_1 \approx L_2\approx L_3 \approx L$) as  
\begin{equation}
  X^{\synthLISA} = -\frac{4 L^2}{c^2} \,\tens{d}^{X} : \ddot\tens{h}\,,\qquad
  X^{\LISAsim} = -\frac{2 L}{c} \, \tens{d}^{X} : \dot\tens{h}\,,
  \label{eq:9}
\end{equation}
where $:$ denotes the contraction of both tensor indices, and 
$\tens{d}^{X}\equiv(\vec{n}_2\otimes\vec{n}_2 - \vec{n}_3\otimes\vec{n}_3)/2$
is the usual LWL response tensor for a GW interferometer with arms
$\vec{n}_2$ and $\vec{n}_3$. The analogous expressions for $Y$ and $Z$ are
obtained by cyclic permutations of the indices 
$1\rightarrow 2\rightarrow 3\rightarrow 1$.   
We define an associated scalar ``strain'' for each of the
detectors $I = X, Y, Z$ as 
\begin{equation}
  \label{eq:11}
  h^{I}(t) \equiv \tens{d}^{I}(t) : \tens{h}\left(\tau(t)\right)\,.
\end{equation}
The timing relation $\tau(t)$ accounts for the Doppler effect
caused by the orbital motion of the detector, namely
$\tau(t) = t - \vec{r}\cdot\vec{e}_z / c\,$, where $\vec{r}(t)$ is
the position of the detector with respect to the SSB, and $\vec{e}_z$
is the propagation direction of the GW. Note that in the LWL
approximation, we can assume that all virtual detectors follow the
same trajectory $\vec{r}(t)$ corresponding to the 
barycenter of the three spacecraft.  

The input to our search code consists of Fourier-transformed data
stretches of duration $T_\SFT$, the so-called SFTs (for Short Fourier Transforms), 
which is a common data format used within the LIGO Scientific Collaboration for
continuous-wave searches (e.g.\ see \cite{lsc06:_coher_scorp_x}).
The time baseline $T_\SFT$ has to be chosen sufficiently short such
that the noise-floor can be approximated as stationary and the
rotation and acceleration of the detector can be neglected.  For LISA
we chose $T_\SFT = 7$\,days, while in LIGO/GEO (where the rotation of
the Earth dominates the acceleration) this is typically $T_\SFT = 30$\,min.
Approximating the detector tensor $\tens{d}^{I}$ as constant during
$T_\SFT$, we can Fourier-transform \eref{eq:9} to obtain   
\begin{equation}
  \hspace*{-1cm}
  \wt{h}^{X}(f) = \frac{1}{(4\pi f L/c)^2}\wt{X}^{\synthLISA}(f)\,,\quad
  \wt{h}^{X}(f) = \frac{i}{4\pi f L/c}\wt{X}^{\LISAsim}(f)\,.
  \label{eq:10}
\end{equation}
We use $\wt{h}^I(f)$ as our SFT input data, and so we can run the same
pipeline on data from LISA~Simulator and synthetic LISA, with only a
different ``calibration'' \eref{eq:10} used to generate the SFTs.
The noise contributions to $X$, $Y$, and $Z$ are correlated, therefore
it is often convenient to work with the TDI variables $X$ and $Y-Z$
instead, which are statistically independent.  This is a
straightforward generalization, using the response tensor
$\tens{d}^{Y-Z} = \tens{d}^{Y} - \tens{d}^{Z}\,$.
Note that $X$ and $Y-Z$ generally have different noise
levels, but this is properly taken into account in the multi-detector
$\F$-statistic.

\subsection{The $\F$-Statistic Method}

The $\F$-statistic was originally developed in \cite{jks98:_data} and
extended to the multi-detector case in \cite{cutler05:_gen_fstat}. 
A generalization to the full TDI framework for LISA was obtained in
\cite{krolak04:_optim_lisa}, but here we follow the simpler route of
working in the LWL approximation, which allows for a more direct
application of existing LIGO/GEO codes to LISA data analysis. 

Combining the scalar strain \eref{eq:11} with the expression
\eref{eq:4} for the GW tensor, we can write the strain signal $h^I$
at detector $I$ as 
\begin{equation}
  h^{I}(t) = \sum_{\mu=1}^4 \mc{A}^\mu\, h^{I}_\mu(t) \,,
\end{equation}
in terms of the four basis functions
\begin{equation}
  \eqalign{
    h^{I}_1(t) = a^{I}(t)\, \cos \phi\left(\tau(t)\right),
    &\qquad
    h^{I}_2(t) = b^{I}(t)\, \cos\phi\left(\tau(t)\right), 
    \\
    h^{I}_3(t) = a^{I}(t)\, \sin\phi\left(\tau(t)\right),
    &\qquad
    h^{I}_4(t) = b^{I}(t)\, \sin\phi\left(\tau(t)\right)\,,
    }\label{eq:8}
\end{equation}
where we defined the antenna-pattern functions 
$a^{I} \equiv \tens{d}^{I}:\tens{\varepsilon}_{\!+}$ and 
$b^{I} \equiv \tens{d}^{I}:\tens{\varepsilon}_{\!\times}$. 
The functions $\{h^I_\mu\}$ depend on the sky-position
$\skypos$ and the frequency $f(\tau)$ of the source. 
We see that the signal parameters separate into two classes: 
(i) the four \emph{amplitude parameters}
$\A \equiv \{\A^\mu\}$ given in (\ref{eq:5}) and (ii) the
\emph{Doppler parameters} 
$\doppler \equiv \{\lat, \lon, f, \dot{f}, \ddot{f}, ... \}$.
We model the output $x^I(t)$ of detector $I$ as a superposition of
stationary Gaussian noise $n^I(t)$ and a signal $h^I(t; \A, \doppler)$.  
Following the notation of \cite{cutler05:_gen_fstat,prix06:_searc}, we 
write the different data-streams $x^I(t)$ as a vector $\mbf{x}(t)$, 
and we define the standard multi-detector scalar product as
\begin{equation}
  \label{e:innprod}
  (\mbf{x}|\mbf{y}) = \sum_{I,J} \int_{-\infty}^{\infty} 
  \wt{x}^{I*}(f)\, S^{-1}_{IJ}(f)\, \wt{y}^J(f)\, df\,,
\end{equation}
where $\wt{x}$ is the Fourier-transform, $x^*$ denotes  complex
conjugation, and $\{S^{-1}_{IJ}(f)\}$ are the elements of
the inverse of the noise-power matrix.
%%\begin{equation}
%%  \label{eq:1}
%%  P^{IJ}(f) \equiv \wt{\kappa}^{IJ}(f)\,,\quad\mbox{with}\quad
%%  \kappa^{IJ}(t) \equiv \expected{n^I(t)\, n^J(0)}\,,
%% \end{equation}
We search for a signal by seeking the parameters $\{\A, \doppler\}$
which maximize the log-likelihood ratio 
\begin{equation}
  \hspace*{-1cm}
  L(\mbf{x}; \A, \doppler ) 
  = (\mbf{x}|\mbf{h}) - \frac{1}{2}(\mbf{h}|\mbf{h})
  = \mc{A}^\mu (\mbf{x}|\mbf{h}_\mu) 
  - \frac{1}{2}\mc{A}^\mu (\mbf{h}_\mu|\mbf{h}_\nu) \mc{A}^\nu\,,
\end{equation}
with automatic summation over
repeated amplitude indices $\mu,\nu$. Defining 
\begin{equation}
  x_\mu(\doppler) \equiv (\mbf{x}|\mbf{h}_\mu)\,,\quad\mbox{and}\quad
  \mc{M}_{\mu\nu}(\doppler) \equiv (\mbf{h}_\mu|\mbf{h}_\nu)\,,
  \label{e:Ametric}
\end{equation}
we see that $L$ is maximized for given $\doppler$ by the amplitude
estimator $\mc{A}^\mu_\cand = \mc{M}^{\mu\nu}x_\nu$, where
$\mc{M}^{\mu\nu}$ 
is the inverse matrix of $\mc{M}_{\mu\nu}$. Thus the detection
statistic $L$, maximized over the amplitude parameters $\A$, is 
\begin{equation}
  \F(\mbf{x}; \doppler) \equiv \frac{1}{2} \, x_\mu \, \mc{M}^{\mu\nu} \,x_\nu\,,
  \label{eq:2}
\end{equation}
which defines the (multi-detector) $\F$-statistic.

\subsection{Analysis Pipeline}

Our analysis was based on standard LAL/LALApps software~\cite{lalapps}
developed for the search for continuous GWs with ground-based
detectors, in particular the code \texttt{ComputeFStatistic\_v2}, which
implements the multi-detector $\F$-statistic \eref{eq:2}. 
Only minor modifications were necessary to adapt this code
to the analysis of LISA data using the LWL approximation.   
All white-dwarf binary signals in the first MLDC had constant
intrinsic frequency $f$, so the set of Doppler parameters to search over
consisted of $\doppler = \{\lat, \lon, f\}$. 
We performed a \emph{hierarchical search} that first runs single-detector
searches on each of the TDI variables $I$, looks for co\"{\i}ncident
local maxima of $2\F$, and in the last step performs a multi-detector
$\F$-statistic search to establish the parameters of each candidate signal.  
Our initial analysis submitted as an MLDC entry~\cite{GWDAW_MLDC1} used
the TDI-variables $X$, $Y$ and $Z$ as three ``detectors'', assuming
for simplicity that their correlation matrix $S_{IJ}(f)$ is
diagonal. However, given that the corresponding noises are correlated,
we subsequently re-ran the search using the \emph{uncorrelated} TDI
variables $X$ and $Y-Z$, which was used for the results presented here
(but did not result in any significant changes in the results).
Whether $I$ ranges through $\{X,Y,Z\}$ or $\{X,\,Y-Z\}$, the structure
of the pipeline is the same: 
\begin{enumerate}

\item Perform a wide-parameter $\F$-statistic search on each data
  stream $I$ over a template grid of Doppler parameters $\{\lat,\lon, f\}\,$. 
  The grid was chosen as isotropic in the sky, with angular mesh size 
  $d \alpha = \sqrt{2 m} / (2\pi\,f\,R_\mathrm{orb}/c)$, with the
  orbital radius $R_\mathrm{orb} = 1\,$AU and we a mismatch of $m=0.3$. 
  The frequency spacing used is
  $d f = \sqrt{12 \,m} / ( \pi \, T)$, where $T = 1\,$y is the
  observation time. 
  These step sizes were computed from the orbital metric \cite{prix06:_searc}.

\item Keep only candidates which are \emph{local maxima} of $2\F$
  (above some threshold), and which are \emph{co\"{\i}ncident} with
  consistent Doppler parameters in all detectors $I$.  
  %(within a ``sphere'' of radius 2 grid spacings)

\item Perform a more finely-gridded multi-detector search around each
  candidate to increase the accuracy of the parameter estimation.

\item Classify each candidate as \emph{primary} if it has the highest
  $2\F$ value within \mbox{$\Delta f = 1.4\times10^{-4}\,f$}, and
  as \emph{secondary} otherwise. 
\end{enumerate}
The last step arises from the empirical observation that a given
signal will have secondary ``false'' $\F$-statistic maxima at
frequencies within roughly $\sim 10^{-4}\,f$ but at different sky
positions.
Only primary candidates were reported, while the secondary candidates
were discarded. %
This is a limitation of our pipeline: given two signals very
close in frequency but at different sky positions, it cannot
distinguish the peak at the true sky position of the ``fainter''
source from a secondary maximum of the ``brighter'' one. This problem
is seen particularly in Challenges 1.1.4 and 1.1.5 with signals
clustered very densely in frequency. 

\section{Results}

\subsection{Challenge 1.1.1: Isolated Binaries}

This challenge consisted of three separate data sets, each containing
one WD signal at an unspecified sky position and within a given
frequency band: in 1.1.1a at $\sim 1\,\mHz$, in 1.1.1b at $\sim 3\,\mHz$, 
and in 1.1.1c at $\sim 10\,\mHz$. 
Note that the LWL is only a good approximation for $f\ll10\,\mHz$, and
we therefore expect it to deteriorate significantly in  1.1.1b 
and 1.1.1c. 
\begin{table}[htbp]
  %%\centering
  \caption{Recovery of Doppler parameters in Challenge 1.1.1:
    $\Delta f$ is the frequency error, and $\phi_{\mathrm{sky}}$ is
    the angle between recovered and true sky position.
    Frequency $f$ and sky position $\skypos$ were accurately
    determined even at the highest frequencies.
  }  
  \label{tab:Challenge111_doppler}
  \begin{indented}
  \item[]\begin{tabular}{@{}l ccccc}
\br
Challenge & $f$ & $\lat$ & $\lon$ & $\Delta f$ & $\phi_{\mathrm{sky}}$ \\
\mr
1.1.1a & $1.1\un{mHz}$ & $0.95\un{rad}$ & $5.07\un{rad}$ & $1.7\un{nHz}$ & $34.8\un{mrad}$ \\
1.1.1b & $3.0\un{mHz}$ & $-0.09\un{rad}$ & $4.63\un{rad}$ & $0.8\un{nHz}$ & $7.1\un{mrad}$ \\
1.1.1c & $10.6\un{mHz}$ & $-0.11\un{rad}$ & $4.66\un{rad}$ & $0.2\un{nHz}$ & $4.4\un{mrad}$ \\
\br
\end{tabular}

  \end{indented}
\end{table}
Nevertheless, in each of the three cases our pipeline recovered a
single primary candidate, and the Doppler parameters were determined
with very good accuracy, as summarized in \tref{tab:Challenge111_doppler}. 
The apparent improvement in the Doppler accuracy seen in
\tref{tab:Challenge111_doppler} is due to statistical
fluctuations. Running this search on a larger number of sources (such
as in Challenge 1.1.2) reveals no clear trends. 
\begin{figure}[htbp]
  \centering
  \mbox{
    \includegraphics[width=0.33\textwidth]{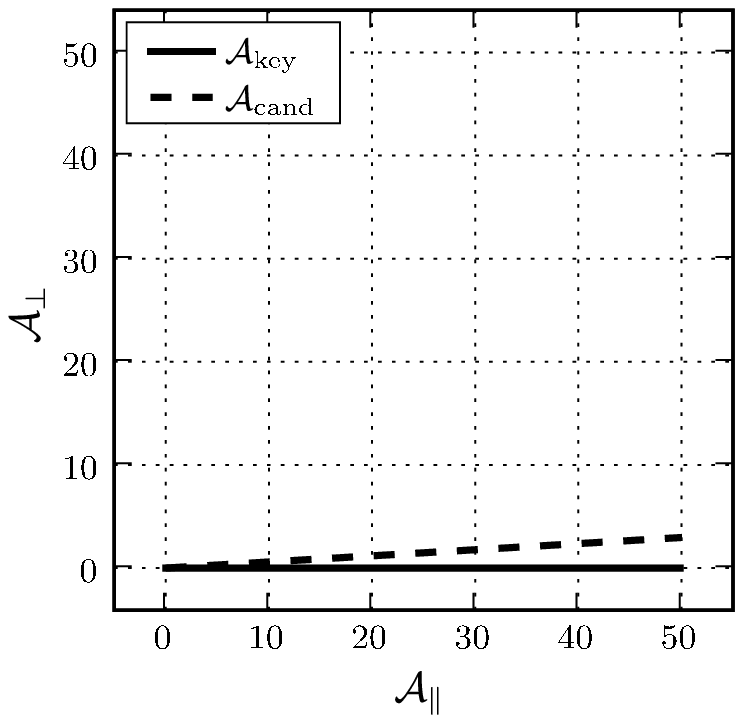}
    \includegraphics[width=0.33\textwidth]{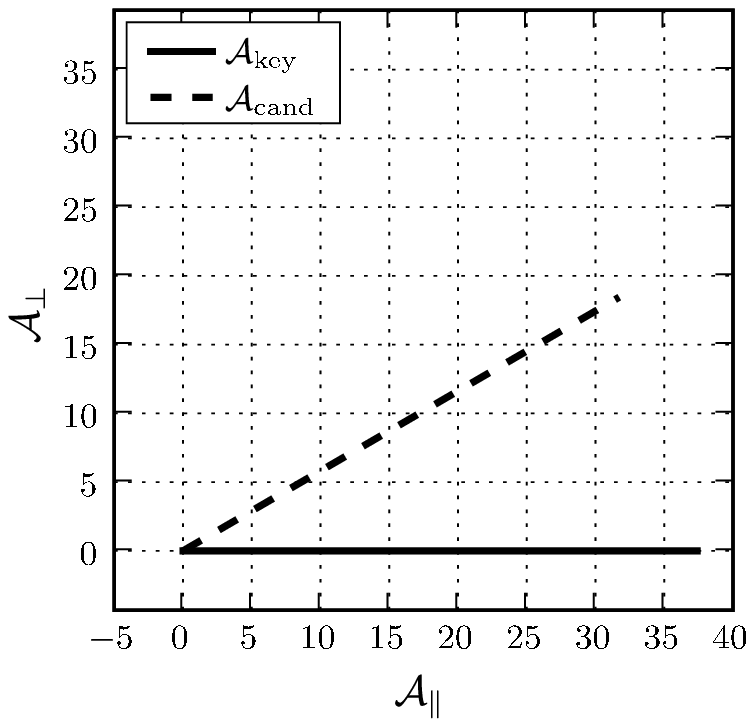}
    \includegraphics[width=0.33\textwidth]{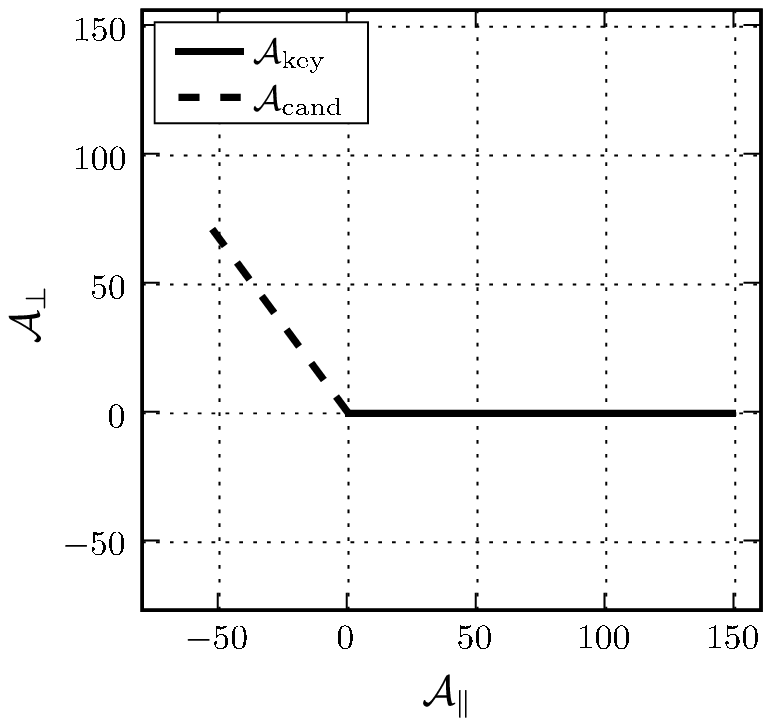}
    }
    \caption{Recovery of amplitude parameters in Challenges 1.1.1a (\emph{left}),
      1.1.1b (\emph{middle}), and 1.1.1c (\emph{right}).
      Each plot compares the recovered amplitude 4-vector $\A_\cand$
      to the injected signal 4-vector $\A_\key$, shown in the plane
      defined by the two vectors.  
      Gaussian fluctuations would lead to a separation of the
      endpoints of the order $\abs{\Delta\A}\sim 2$.
      The breakdown of the LWL with increasing frequency leads to
      larger errors, affecting both the orientation and the magnitude
      of the recovered amplitude vector.}  
  \label{fig:Challenge111_amp}
\end{figure}
The recovery of the amplitude parameters $\A$ is illustrated in
\fref{fig:Challenge111_amp}, comparing the estimated 4-vector $\A_\cand$
to the 4-vector $\A_\key$ of the injected parameters. The amplitude
4-vectors $\A$ live in a space with constant metric tensor $\M_{\mu\nu}$ given in
\eref{e:Ametric}, so the norm is 
$\abs{\A}^2 \equiv \A^\mu\, \M_{\mu\nu} \, \A^\mu$.
The two vectors $\mc{A}_\cand$ and $\mc{A}_{\key}$ define a plane, and
so we can plot them in two dimensions, with the horizontal and
vertical components  
\begin{equation}
  \mc{A}_{\parallel} = 
  \frac{\mc{A}_{\key}\cdot\mc{A}}{\abs{\mc{A}_{\key}}}\,,
  \qquad
  \mc{A}_{\perp} =
  \abs{
    \mc{A} - \mc{A}_{\parallel}
    \frac{\mc{A}_{\key}}{\abs{\mc{A}_{\key}}}
  }\,,
\end{equation}
where the inner product is calculated using the metric $\M_{\mu\nu}$. 
These components are shown in \fref{fig:Challenge111_amp}, and we
see that the agreement of the amplitude parameters deteriorates
substantially for higher frequencies, where the LWL approximation
breaks down.

If the deviation is caused by noise \emph{alone}, then the difference
$\Delta\mc{A} \equiv \A_\cand - \A_\key$ between the amplitude vectors
has zero mean, i.e., $\expected{\Delta\A} = 0$, and covariance 
$E[\Delta\A^\mu \, \Delta\A^\nu] = \M^{\mu\nu}$, 
where $\expected{\ldots}$ denotes the expectation value.
The magnitude of this difference, 
$\abs{\Delta\A} = \sqrt{ \Delta\A^\mu \, \M_{\mu\nu} \, \Delta\A^\nu}$,
would have variance $E[\abs{\Delta\A}^2] =\M_{\mu\nu}\M^{\mu\nu} = 4$. 
Therefore $\abs{\Delta\A}/2$ measures the difference between the two
amplitude vectors in terms of the number of standard deviations.
It is also instructive to compare the \emph{magnitude} of the
recovered versus the injected amplitude vector.
The magnitude $\abs{\A_\key}$ of the injected signal is equivalent to
the optimal signal-to-noise ratio (SNR). 
Note, however, that $\abs{\A_\cand}^2$ is a \emph{biased} estimator
for $\abs{\A_\key}^2$, namely
\begin{equation}
  \label{eq:3}
  \expected{ \abs{\A_\cand}^2 } = 
  \expected{ 2\F } = 4 + \abs{\A_\key}^2\,.
\end{equation}
Therefore we use the following measure for the error in the norm of
the recovered amplitude vector $\A_\cand$:
\begin{equation}
  \label{eq:13}
  \epsA \equiv \frac{\abs{\A_\cand}^2 - \abs{\A_\key}^2 - 4}{2 \, \abs{\A_\key}^2}\,,
\end{equation}
which is unbiased, i.e., $\expected{\epsA} = 0$.
The standard deviation of $\abs{\A_\cand}^2$ is 
$2(2 + \abs{\A_\key}^2)^{1/2} \approx 2 \abs{\A_\key}$, 
and so the expected error $\epsA$ from noise alone
would be $\expected{\epsA} \approx \abs{\A_\key}^{-1}$.
\begin{table}[htbp]
  \caption{Errors in recovered amplitude parameters in Challenge 1.1.1: 
    as seen in \fref{fig:Challenge111_amp}, the angle $\phi_{\A}$
    between $\A_\cand$ and $\A_\key$ grows with increasing frequency,
    and there is an increasing deficit in the magnitude
    $\abs{\A_\cand}$ with respect to the SNR $\abs{\A_\key}$,
    as quantified by $\epsA$. 
    The absolute error in $\abs{\Delta\A}/2$ from Gaussian noise would
    be expected to be $\sim \O(1)$, while for $\epsA$ and $\phi_\A$ it
    would be $\abs{\A_\key}^{-1}$.
  }
  \label{tab:Challenge111_amp}
  %\centering
  \begin{indented}
    \item[]\begin{tabular}{@{}l ccccc}
\br
Challenge & $f$ & $\abs{\A_{\key}}^{-1}$ & $\epsA$ & $\phi_{\A}$ & $\abs{\Delta\A}/2$ \\
\mr
1.1.1a & $1.1\un{mHz}$ & $0.020$& $0.005$& $0.059$& $1.5$\\
1.1.1b & $3.0\un{mHz}$ & $0.027$& $-0.020$& $0.527$& $9.6$\\
1.1.1c & $10.6\un{mHz}$ & $0.007$& $-0.306$& $2.207$& $108.7$\\
\br
\end{tabular}

    \end{indented}
\end{table}
\Tref{tab:Challenge111_amp} summarizes the errors in the amplitude
parameters for the three challenge data sets in terms of 
$\abs{\Delta\A}/2$, the relative difference $\epsA$ of the norms, and
the angle $\phi_\A$ between the recovered and the injected amplitude
vectors, given by   
\begin{equation}
  \phi_{\mc{A}}
  =
  \cos^{-1}
  \left(
    \frac{\mc{A}_{\cand}\cdot\mc{A}_{\key}}
    {\abs{\mc{A}_{\cand}}\abs{\mc{A}_{\key}}}
  \right)
  \,.
\end{equation}
We see in \tref{tab:Challenge111_amp} that the amplitude errors are
larger than would be expected from noise fluctuations alone,
especially at higher frequencies, which is consistent with the
breakdown of the LWL.

\subsection{Challenge 1.1.2: Verification Binaries}

In Challenge 1.1.2, the sky position and
frequency of twenty ``verification binaries'' was given, 
while the amplitude parameters of the injected signals were unknown.
We therefore performed a targeted $\mc{F}$-statistic search
at each of the specified sets of Doppler parameters, and found the
maximum-likelihood estimators $\A_\cand$ for the amplitude parameters.
\begin{figure}[htbp]
  \mbox{
    \hspace*{-0.8cm}
    \includegraphics[width=0.36\textheight]{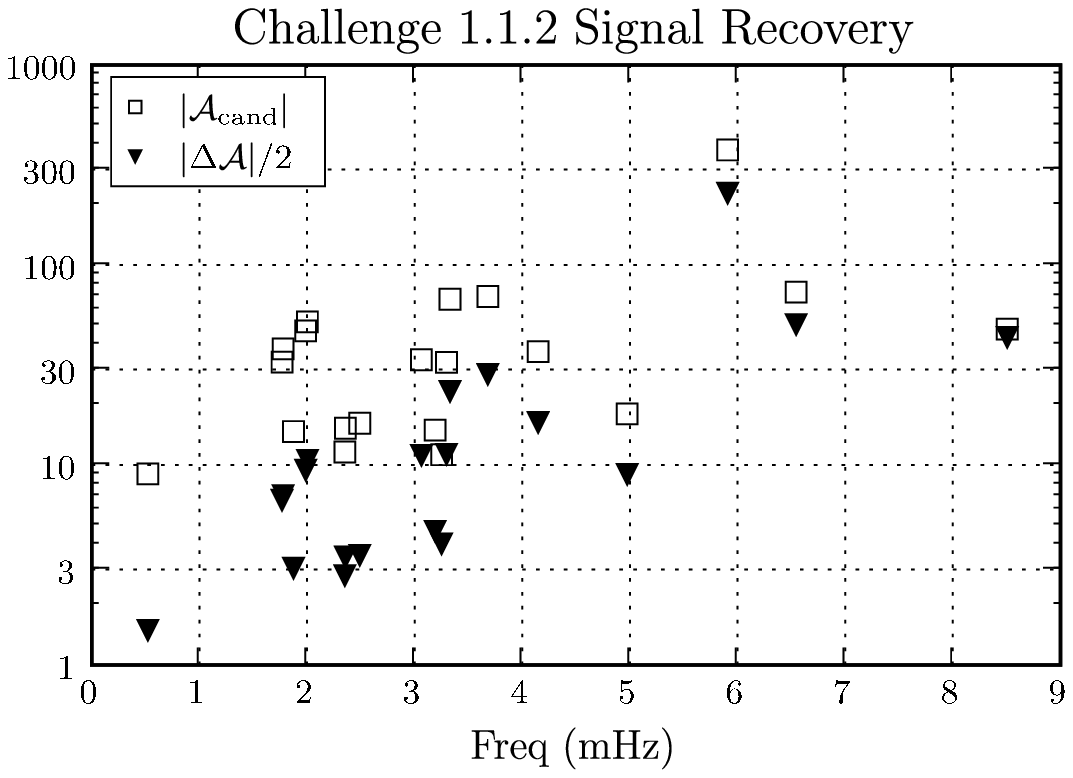}
    \hspace*{-1.0cm}
    \includegraphics[width=0.36\textheight]{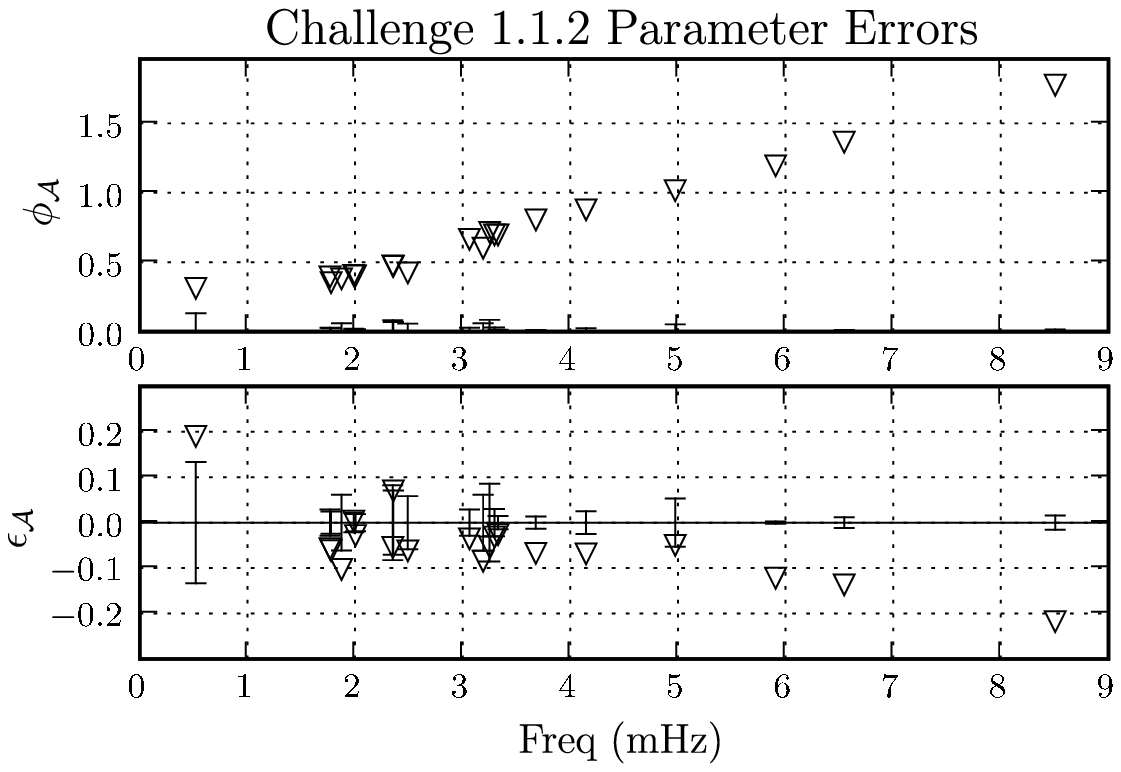}
  }
  \caption{Recovery of amplitude parameters in Challenge 1.1.2.
    \emph{Left:} all 20 signals are
    recovered with $\abs{\A_\cand} \ge 8.5$, but the errors 
    $\abs{\Delta\A}/2$ are substantially larger than the expected
    standard deviation of unity for all but the smallest frequencies.  
    \emph{Top right:} the angle $\phi_{\A}$ between the true and
    recovered amplitude vectors grows with frequency, and is always
    larger than its expected standard deviation of $\abs{\A_\key}^{-1}$.  
    \emph{Bottom right:} the norm of the recovered amplitude vector
    is within the expected range of $\abs{\A_\key}^{-1}$ for much of
    the frequency band, but begins to show a deficit for $f > 5\,\mHz$. 
  }
  \label{fig:Challenge112}
\end{figure}
\Fref{fig:Challenge112} illustrates the discrepancies between the 
recovered $\A_\cand$ and the injected amplitude parameters $\A_\key$,
in terms of $\abs{\Delta\A}$,  $\epsA$, and $\phi_{\A}$.
Again we see that our recovered amplitude parameters differ from the
injected ones by more than would be expected from Gaussian noise
alone, and that the agreement deteriorates at higher frequencies.

\subsection{Challenge 1.1.3: Resolvable Binaries}

Challenge 1.1.3 was a blind search on data containing 20 white dwarf
binary signals across the LISA band. 
\begin{figure}[htbp]
  \mbox{
    \hspace*{-0.8cm}
    \includegraphics[width=0.55\textwidth]{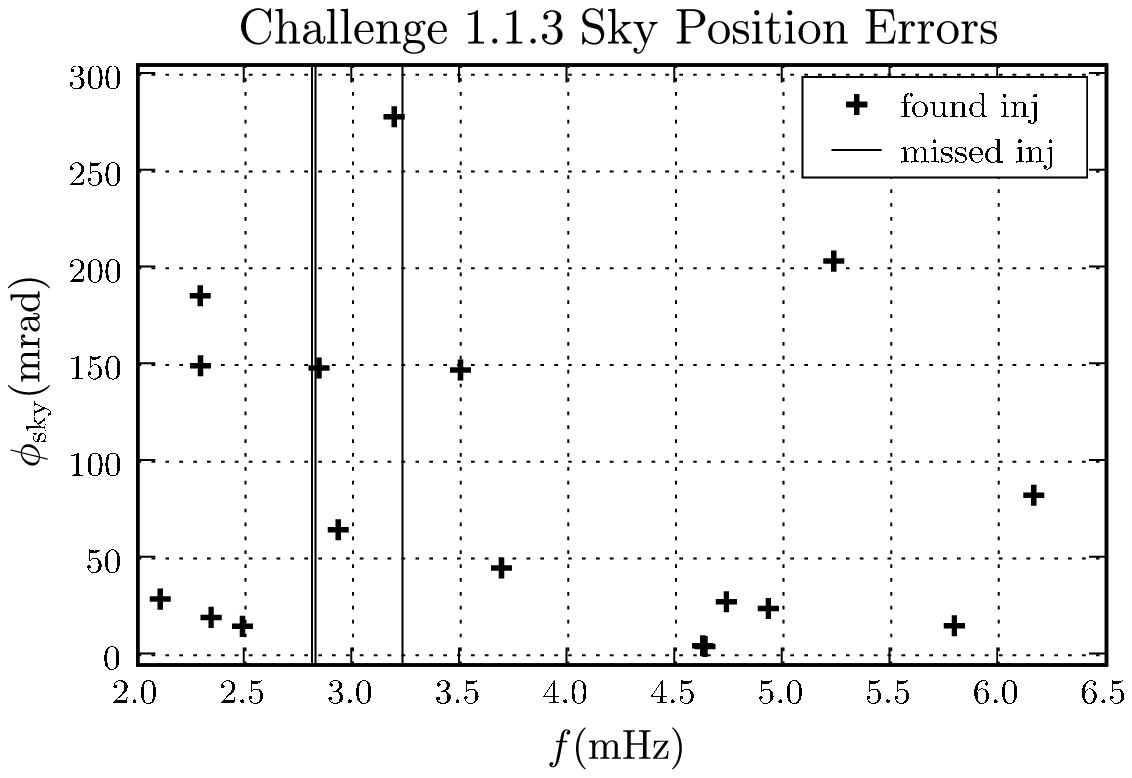}
    \hspace*{-0.5cm}
    \includegraphics[width=0.55\textwidth]{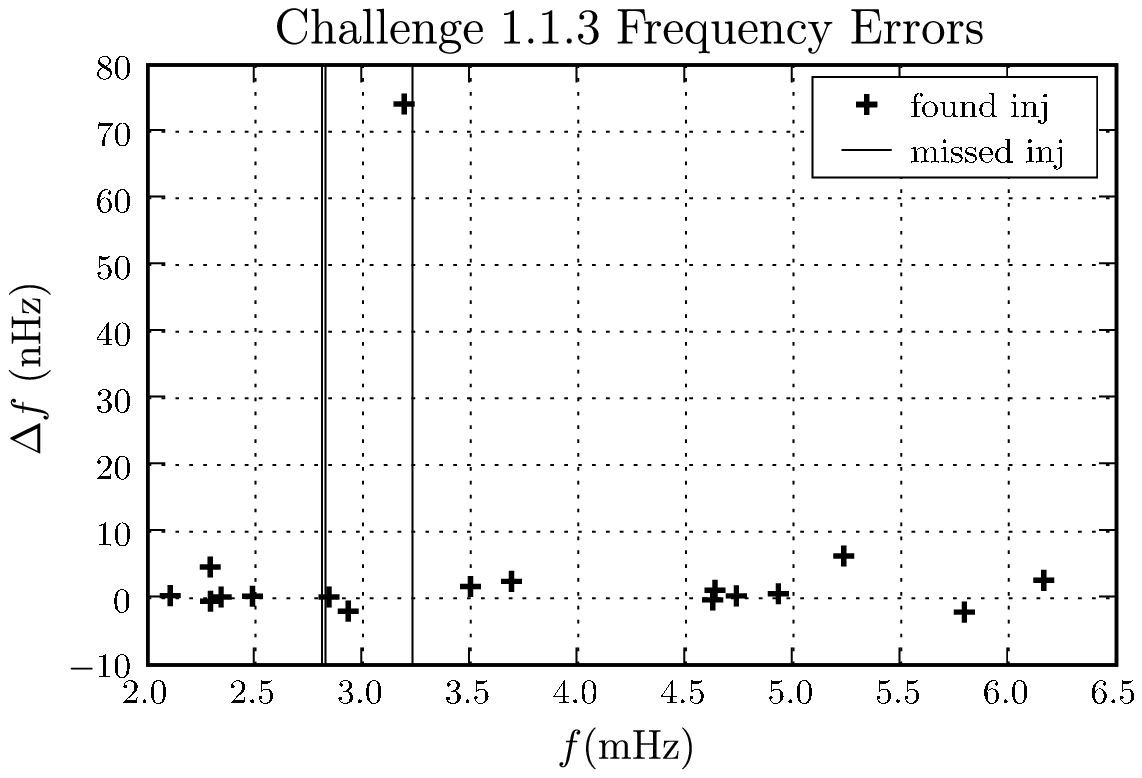}
    }
  \caption{Doppler parameter recovery in Challenge 1.1.3: errors in sky
    position (\emph{left figure}) and frequency (\emph{right figure})
    as functions of frequency.
    The width of the frequency co\"{\i}ncidence window, i.e.,
    $1.4\times10^{-4}f$, is too small to be seen on this scale,
    so no error bars are shown on the found injections.
    The three missed signals (long vertical lines) all fall close to
    recovered signals, but outside of all co\"{\i}ncidence windows.}
  \label{fig:Challenge113}
\end{figure}
As shown in \fref{fig:Challenge113}, we recovered 17 of the 20 signals
with good frequency and sky accuracy. The three missed signals were at
frequencies ``close'' to recovered sources, but not within the 
frequency co\"{\i}ncidence window of $1.4\times10^{-4}f$, and there is
some indication that the Doppler parameters of those sources were
slightly compromised.  

\subsection{Challenge 1.1.4 and 1.1.5: Source Confusion}

In Challenges 1.1.4 and 1.1.5, many sources were injected into a small
frequency band in order to illustrate the source confusion problem, 
namely 45 signals within $[3,\,3.015]\,\mHz$ in Challenge 1.1.4 and
33 signals within $[2.9985,\, 3.0015]\,\mHz$ in Challenge 1.1.5.
\begin{figure}[htbp]
  \mbox{
    \hspace*{-0.8cm}
    \includegraphics[width=0.55\textwidth]{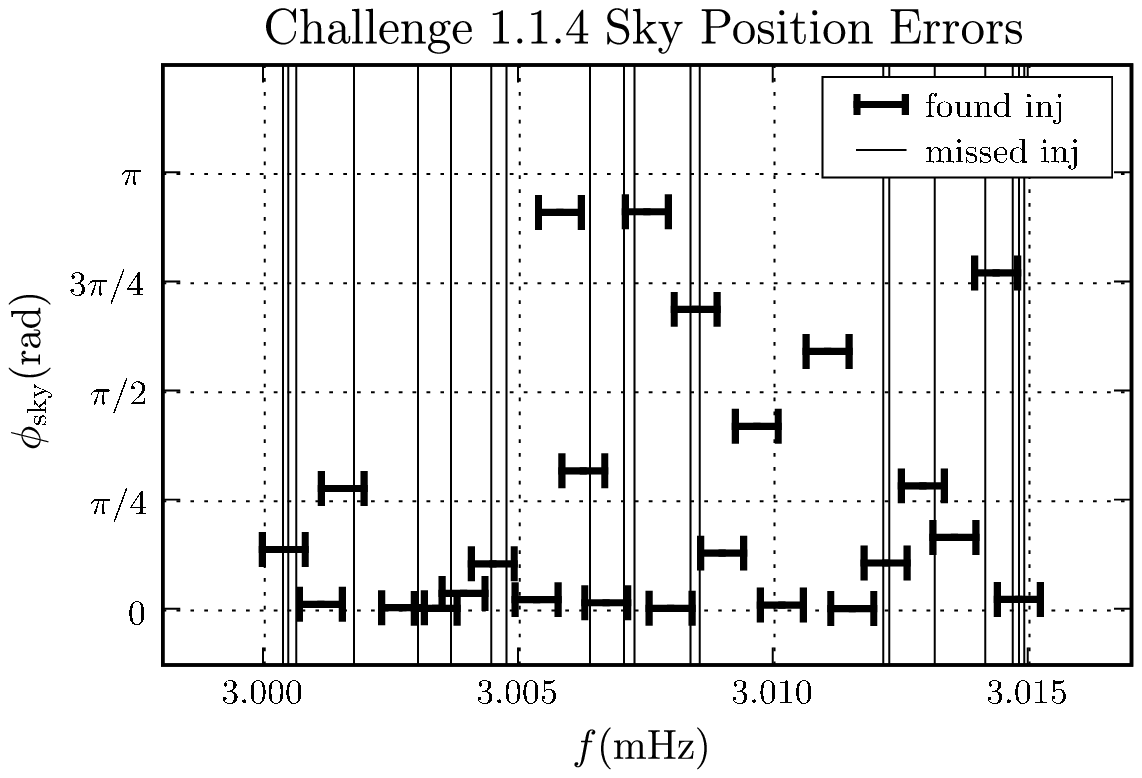}
    \hspace*{-0.5cm}
    \includegraphics[width=0.55\textwidth]{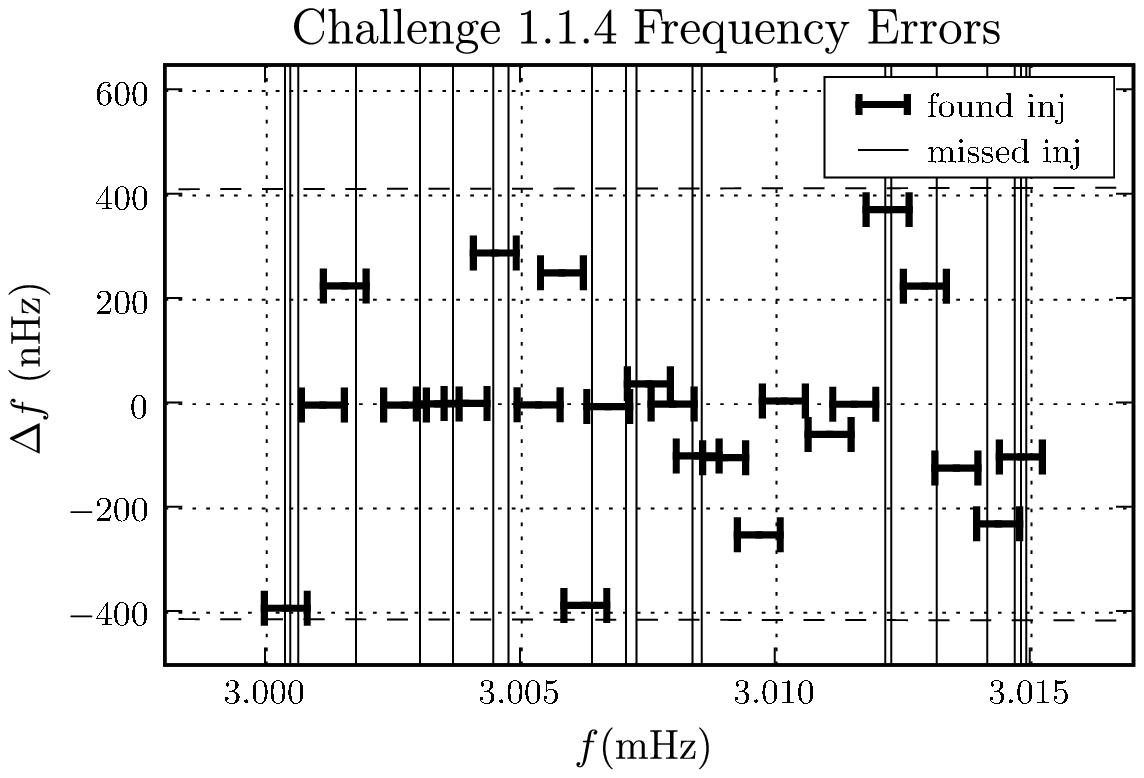}
    }\\[0.5cm]
    \mbox{
      \hspace*{-0.8cm}
      \includegraphics[width=0.55\textwidth]{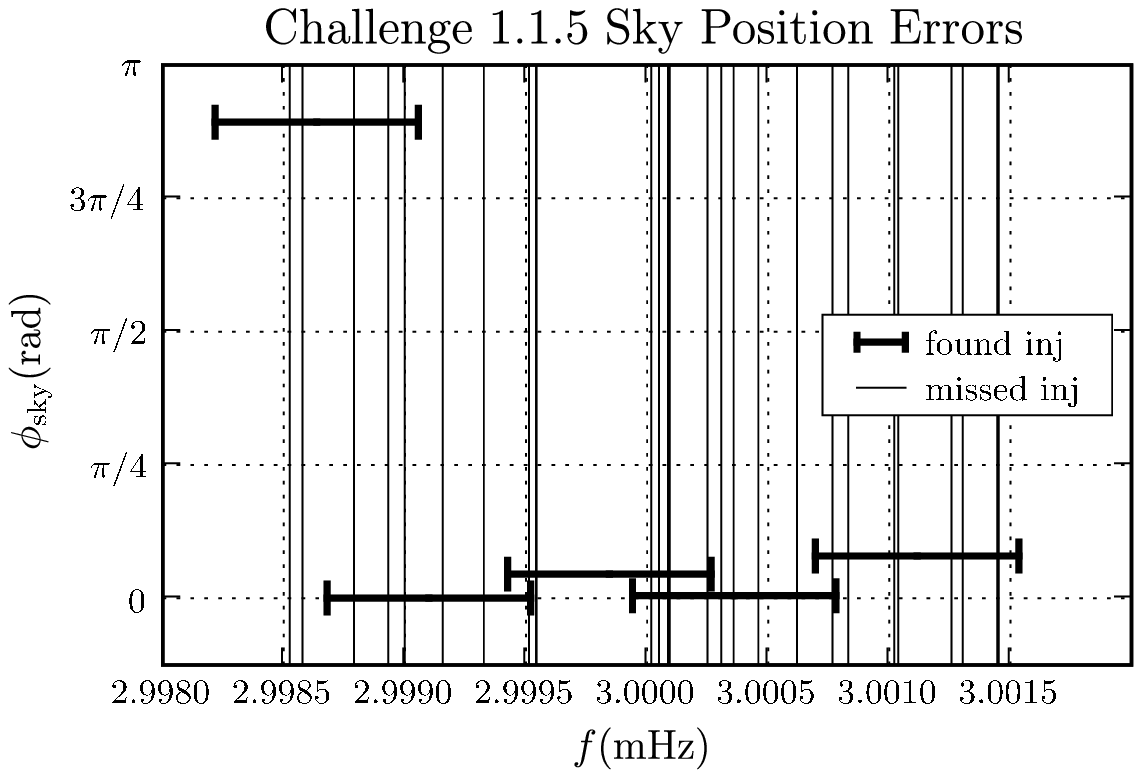}
      \hspace*{-0.5cm}
      \includegraphics[width=0.55\textwidth]{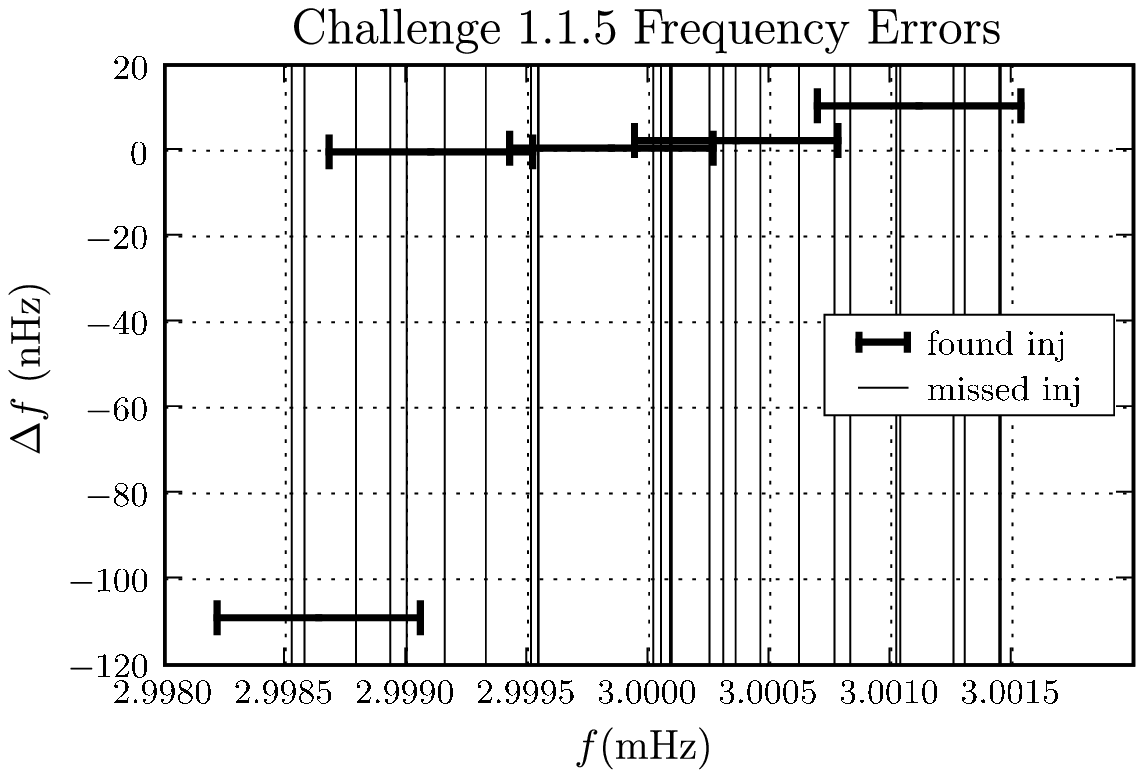}
      }
  \caption{Doppler parameter recovery in Challenges 1.1.4 (\emph{Top row})
    and 1.1.5 (\emph{bottom row}): errors in sky position (\emph{left column}) 
    and frequency (\emph{right column}) as functions of frequency.
    Each of the ``missed'' injections falls within the
    co\"{\i}ncidence window of a recovered signal, and would therefore
    have been rejected as a secondary maximum.  The dashed lines in
    the top-right plot show the maximum possible frequency recovery
    error, namely the width of the co\"{\i}ncidence window. 
    In both challenges, source confusion causes our pipeline to find
    a candidate at every possible frequency, including one false
    alarm at $f=3.0022\mHz$ in Challenge 1.1.4.}
  \label{fig:Challenge114}
\end{figure}
As shown in Figure~\ref{fig:Challenge114},
our pipeline ``found'' signals all across the band, namely 25 signals
in Challenge 1.1.4 and only 5 signals in Challenge 1.1.5, but many
of them were far removed in sky position from any true signal.  Many
additional signals were missed within the frequency co\"{\i}ncidence
window, presumably because they were mistaken for secondary maxima of
the ``found'' signals.
The results of this challenge illustrate a known limitation of the
pipeline used here: it cannot distinguish multiple signals too close
together in frequency.

\section{Conclusions}

Using the $\F$-statistic in the long-wavelength limit approximation,
we found that the estimation of the four \emph{amplitude parameters}
$\{\A^\mu\}$  deteriorates significantly with increasing frequency, as
would be expected from the breakdown of the LWL.
However, the \emph{detection} of signals and the
estimation of the \emph{Doppler parameters} (frequency and
sky-position) does not seem to be affected by the use of the LWL, even
at frequencies as high as $f \sim 10\,\mHz$.  
This somewhat surprising result suggests the following
``hierarchical'' search strategy: start with a fast $\F$-statistic
code using the LWL to detect signals and localize them in Doppler
space, then use a more accurate (and computationally expensive)
modelling of the TDI responses to estimate the amplitude parameters.

We are planning to study these findings in a more systematic way using
larger number of signals. 
More work is required to deal with ``source confusion'', i.e.,
signals that lie within a frequency window $\O(10^{-4}\,f)$.
Secondary maxima in parameter space due to a signal cannot easily
be distinguished from primary maxima corresponding to other signals
within this frequency window. One popular strategy consists of
successively ``removing'' detected signals from the data, which also
eliminates its associated secondary maxima, and allows one to re-run
the search for the next-loudest candidates. An alternative approach might
consist of a classification of candidates into \emph{equivalence classes} 
consistent with the same signal, either by using the metric
or a suitable global correlation criterion analogous to the ``circles
in the sky'' \cite{prix05:_circles_sky} present for short
observation times. 

\ack

We thank Stas Babak for a crash course in TDI and Curt Cutler for
helpful discussions.    
This work was supported by the Max-Planck-Society.
This paper has been assigned LIGO Document Number \dcc.

\section*{References}
\bibliography{biblio}

\end{document}